\input{aipcheck}

\documentclass[
    ,final            
  ]
  {aipproc}

\layoutstyle{6x9}
\usepackage{wrapfig}

  \usepackage{multirow}

\begin{document}

\title{An alternative singularity-free cosmological scenario from cusp geometries}

\classification{98.80.-k, 90.80.Bp, 47.20.k, 02.40.Xx, 02.40.Dr}
 
 
 


\keywords      { Physics of the early universe; Spacetime metrics; Friedmann Singularity-free curvature}

\author{Reinaldo R. Rosa}{
  address={Lab for Computing and Applied Math (LAC) - National Institute for Space Research (INPE),\\ S\~ao Jos\'e dos Campos, SP, Brazil}
  }
\author{Cristiano Strieder}{
  address={Lab for Computing and Applied Math (LAC) - National Institute for Space Research (INPE),\\ S\~ao Jos\'e dos Campos, SP, Brazil}
}
\author{Diego H. Stalder}{
  address={Lab for Computing and Applied Math (LAC) - National Institute for Space Research (INPE),\\ S\~ao Jos\'e dos Campos, SP, Brazil}
}

\begin{abstract}
We study an alternative geometrical approach on the problem of classical cosmological singularity. It is based on a generalized function $f(x,y)=x^2 + y^2 = (1-z)z^n$ which
consists of a cusped coupled isosurface. Such a geometry is computed and discussed
into the context of Friedmann singularity-free cosmology where a pre-big bang scenario is considered. Assuming that the mechanism of cusp formation is described by non-linear oscillations of a pre-big bang extended very high energy density field ($> 3\times 10^{94} kg/m^{3}$), we
show that the action under the gravitational field follows a tautochrone  of revolution, understood here as the primary projected geometry that alternatively 
replaces  the Friedmann singularity  in the standard big bang theory. As shown here this new approach allows us to interpret 
the nature of both matter and dark energy from first geometric principles.\end{abstract}
							
\maketitle

\section{Introduction}	

The classical Friedmann Singularity-Free (FSF) theories  usually  consider more complex structures than the simple spherically symmetric pointlike of a Friedmann singularity (null dimension with infinite density) \cite{ellis77, szekeres60}. Indeed, despite the high isotropy and homogeneity which have been inferred from recent observations, most FSF approaches have considered anisotropic and inhomogenoeus scenarios in the cosmological past\cite{wiltshire08,  misner69, belin82, berger90}.  In such cases the Universe evolution is described as the chaotic oscillation of a point-like particle in the so-called minisuperspace potential well (see, for example, the Bianchi Model-Type I, whose one of the Kosner solutions leads to a cigar-like 'singularity'). We remark the importance of recent FSF theories assuming an initial discrete spacetime, like quantum-loops \cite{bojowald}.  As far as we know, there is no FSF model assuming a pre-big bang single geometric configuration from what the big bang Universe was emerged remaining connected throughout its evolution. In this sense, we introduce here a new geometric scenario that, for simplicity, incorporates ingredients that alternatively admit a non-bouncing FSF approach where both the nature of dark matter and dark energy are interpreted from geometric first principles. 

Our formalism begins with the projected geometry drawn through general physical principles, most of them compatible with most cosmologies that admit a 4-dimensional expanding spacetime.  The formalism of the cosmological cusp should be introduced to solve the problem of the energy density contrast unresolved by the $\Lambda$CDM model. Therefore, this initial version we are assuming that the mechanism of cusp formation is described by non-linear oscillations of a pre-big bang extended very high energy density field ($> 3\times 10^{94} kg/m^{3}$). In fact, our aim in the present paper is to introduce a new intuitive geometric notion which can be grounded in a more rigorous mathematical and physical framework.

Our starting point here is to address the formation of a primordial cusp geometry from which a coupled classical spacetime could emerges (this picture is illustrated in Fig.1).  Following this path, we have initiated a program known as {\it cusp cosmology} based on the cusp-like geometries resulting from nonlinear wave theory (See Appendix).\footnote{The research program began in 2002 at LAC-INPE, Brazil. However, the first formal presentation of this idea was carried out by the first author in the fundamental physics H.1. section on dark matter and dark energy held at COSPAR 2010 in Bremen, Germany.}

\begin{figure}[h]
     \includegraphics[width=0.15\textwidth]{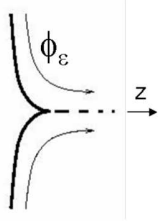}\\
\caption{ A hypothetical projected spacetime (considering $z=ct$) with the energy of the structure flowing($\phi_{\epsilon}$) through a cusp geometry. This picture is called here the energetic cusp structure for which, in a cosmological context, it is conjectured an energy density $> 3\times 10^{94} kg/m^{3}$.}
 \end{figure}

\section{The Projected Cusp Geometry and Its Hamiltonian}

Let us first consider the existence of a projected coupled geometry (hypersurfaces) for the 4-dimensional spacetime $(x,y,z,v\tau)$. 
In such case our object is a 3-dimensional spacetime $(x,y,z=v\tau)$.  Then, in this section 
we look for a set of solutions of polynomial equations (an algebraic variety) that represents two-dimensional surfaces, which must be consistent with the energetic cusp hypothesis presented earlier. Our first inspiration is based on 
the curve ($y=f(x)$) named {\it Conchoid of Nicomedes} (CN) which has the form  $x^2+y^2=b^2x^2/(x-a)^2$. Performing an appropriate representation, one can see that, in the range $-2 < a/b < 0$, there is the formation of the cusp as well as the structure after the cusp when $a/b$ tends to zero. After the CN cusp, when $z>0$, a family of closed structures emerges as for example: right strophoid, lemniscate and projections of isosurfaces as Tschirmhausen cubic and piriforms\cite{Ghomi}. Therefore, extrapolating for $z=f(x,y)$, we focus on the following generalized form:

\begin{equation}
	x^2 + y^2 = (1-z)z^n
\end{equation}

which can be represented parametrically as

\begin{equation}
	x(\varphi, \theta) = a \varphi^{n/2} \sqrt{\frac{2(1-\varphi)}{n}} cos \theta,
\end{equation}

\begin{equation}
	y(\varphi, \theta) = a \varphi^{n/2} \sqrt{\frac{2(1-\varphi)}{n}} sin \theta,
\end{equation}

\begin{equation}
	z(\varphi, \theta) = a \varphi^{n/2}.
\end{equation}

Based on a marching cubes algorithm\cite{petitjean92}, the isosurface can be constructed from a data volume generated by equation 4. Fig.2 shows the output isosurface for $n=4$. Our computing inputs are: (i) a range of $(-1.5, 1.5)$ for $x$ and $y$; (ii) $-1 \leq z \leq 1$; discretized in a $30\times 30\times 30$ grid. The parametric refinement was performed making $a=1$.

\begin{figure}[h]
     \includegraphics[width=0.4\textwidth]{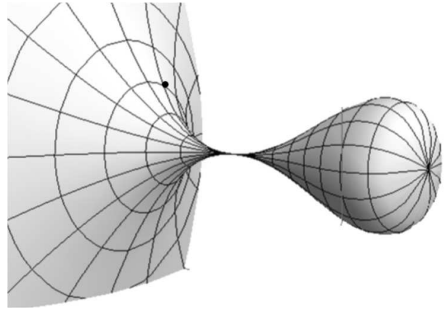}\\
\caption{ The projected cusp spacetime coupling two inhomogenous structures (left: pre-big bang open pure energetic tautochrone of revolution having negative curvature and, right: post-big bang closed low density baryonic space-time.}
 \end{figure}

It is important to mention that, from the parametric representation, coefficients of the first fundamental form
can be obtained and then the line element is given explicitly by the Riemannian metric which
determines the arc length of a curve on a surface. In particular, there exist a family of cubic and quintic surfaces for which the respective Gaussian curvatures have been calculated. Based on such mathematical formula, here a generalized
Gaussian curvature, as a function of $z$ only, can be given implicitly by $K = \frac{n^2}{z^{n/2}}f(n,z)$. Thus, the determination of a geometry designed for the cusp spacetime is reduced to the task of finding the functions f(n, z). Once this is achieved, the isometric surfaces in a Riemannian metric can then be incorporated. 
At least for the case n = 4, the Ricci flow for each element of energy (Tachyon hypothetically) toward the cusp will follow a line of action on a tautochrone of revolution which can be parametrized as

\begin{eqnarray}
z&=&x(\theta)=\frac{a}{2}(\theta-\sin \theta)+x_{L}-\frac{a\pi}{2}\\
\rho &=&y(\theta)=\frac{a}{2}(1+\cos \theta)+y_{L}, \hspace{2mm}  with\hspace{5mm}  0\leq\phi\leq 2\pi
\end{eqnarray}

Note that this propertie support a new perspective of cosmological singularity from where an action can be defined. Considering that 
$x=l(\theta +\sin\theta)$ and $z=l(1 -\cos\theta)$ the tautochrone Hamiltonian and its related action can be written as

\begin{eqnarray}
H=\frac{p_{0}^{2}}{4\,m\,l^{2}(1+\cos \theta)}+m\,g\,l(1 +\cos\theta)\quad \mid\theta\mid<\pi
\end{eqnarray}

\begin{eqnarray}
S(z)=\int_{0}^{l}{\frac{\sqrt{1+z^{\prime}(x)^{2}}}{\sqrt{h-z(x)}}dx}
\end{eqnarray}

where the following conditions are hold: 

$t:[0,l]$, $z\rightarrow  z(x)$, $z(0)= 0$, $z(h)= l$, $(l,h)\rightarrow  (0,0)$. \\

Thus, from the point of view of the physical formalism, the most appropriate model to be adopted for cusp cosmology are those non-homogeneous admitting a quantum gravity approach in the cusp structure, the region where one should consider the baryogenesis. This implies novel approaches from quantum mechanics to General Relativity Theory.  Recently, it has been shown that Einstein equations can emerge as the equation of state of a quantum gravitational system \cite{smolin}. Such cases include  of an  intricate geometric formalism which  can be possibly unified in a simple way such as are possible in the cusp cosmological  scenario we have introduced here.

\section{Concluding Remarks}

In this paper we have concerned with alternative projected geometry for dealing with the problem of Friedmann singularity in physical cosmology. The energetic cusp hypothesis introduced here motivate many interesting questions, which we have only begun to explore. Contrary to the classical version of the Big Bang cosmological model our geometric formalism contains a causal (+,-) structure at the start of  timelike baryonic moment ($t=0$), where all timelike geodesics have a spacelike extension into the tachyonic-like past: a massless "dark energy" negative curvature hypersurface. One of the interesting aspect of this cusped coupled scenario is that both dark energy and dark matter can be addressed into the cosmological apparatus as natural physical ingredients: continuous flux of energy from the warping negative curvature structure and possible spacetime deformations of the secondary positive (or locally flat) curvature structure due to the cusp cosmological constraint (gravity-like stretched spacetime).

It is worthily to note that our approach starts from geometric first principles directly at the level of established geometrical properties. Therefore, the  consistency of the cusp energetic hypothesis introduced here should be verified by existing theories as well as by those in development. In this sense, the second part of this article will aim to provide a comprehensive approach that can select the best candidates for describe the energetic cusp (+.-) in a complete cosmological approach pointing out  possible observational constraints \cite{Li11, Stalder12}.

\section*{Acknowledgments}

The authors acknowledge the financial support from FAPESP and CNPq. RRR is very grateful for (sometimes short but extremely important) discussions with T.B. Veronese, O. Bertolami, M. Makler, D. Bessada, V.A. Belinskii, J. Pontes, D. Bazeia, D. Muller,  F. Abdalla, H.C. Velho, M. Viana, S. Carvalho, F. Salles and A. Froes.

\appendix
\section{Appendix}

{\footnotesize

\subsection{A Possible Physical Cusp Formation}

The freely nonlinear density wave theory allows to describe the density cusp formation from harmonic oscillations [13]. Then, let us discuss one example in attempt to construct a cusp geometry  induced by  parametric instability ocurring for density waves. 

In common with other approaches of the free waves, we adopt a modified (normalized rotation) shearing box approximation (SBA) [14] from which the solutions of the wave equation can be obtained [15,16]. Using an equivalent Lagrangian formalism as introduced by Fromang and Papaloizou [16] to study parametric instabilities in accretion disks, it has been defined a $\phi(\phi_0, \tau)$ to be the position of a energy flux element with $\phi=\phi_0$ in the absence of any instability. If the  component of the momentum conservation equation is defined in all coordinates, the pressure is uniform and the equation of motion in the $\phi$ direction is given by [16].

\begin{equation}
\frac{D^2\phi}{D\tau^2} + (\phi-\phi_0) = - \frac{v^2}{\rho} \frac{\partial \rho}{\partial \phi'}
\end{equation}

where $\rho$ is the energy density and $v$ is the isothermal propagation speed. The derivatives are taken on line elements ($\tau, \phi'$) where $\tau$ is a timelike domain and $\phi'$ is a spacelike domain of the flux direction.

Following Fromang and Papaloizou [16] we can consider the conservation of energy. Then, in order to express $\rho$ in terms of $\phi$ we must have

\begin{equation}
\rho \frac{\partial \phi}{\partial \phi_0} \equiv \rho \left ( 1 + \frac{\partial \xi}{\partial \phi_0} \right ) = \rho_0
\end{equation}

where $\xi(\phi_0,\tau)=\phi(\phi_0,\tau)-\phi_0$. Since the boundary conditions are periodic in SBA, the traveling wave solutions depend only on the phase $\Phi=\phi_0 - U\tau$, such that $\xi=\xi(\Phi)$ with $U$ being the phase velocity. The governing equation that describes traveling wave propagation is then obtained using Eq.(2) in Eq. (1),

\begin{equation}
U^2 \frac{d^2 \xi}{d \Phi^2} + \xi = \frac{v^2}{(1 + \frac{d \xi}{d \Phi})^2} \frac{d^2 \xi}{d \Phi^2}.
\end{equation}

Note that, the first derivative of $\xi$ respect to $\Phi$ can be periodic so that it is possible to interpret the problem as corresponding to the motion of a particle in a potential well given by

\begin{equation}
 V(p)\approx \frac{z^2}{2}(1-(v/U)^2)\hspace{1cm}  where\hspace{1cm} 	z = \frac{d \xi}{d \Phi}.
\end{equation}
 
In this linear limit the system behaves as an harmonic oscillator whose solution is given by $\xi = \xi_0 \cos(k\phi_0-\omega\tau)+\xi_1 \sin(k\phi_0-\omega\tau)$ where $k$ is the wavenumber and $\omega=kU$ is the wave frequency.

Finally, the study of cusp formation can be performed applying a finite difference scheme\cite{haw95} to solve the system of two coupled first order differential equations:

\begin{equation}
	\frac{d \xi}{d \Phi} = z,
\end{equation}
\begin{equation}
	\frac{d z}{d \Phi} = - \left ( \frac{1}{v} \right )^2 \frac{(v / U)^2 \xi}{1 - \frac{(v / U)^2}{(1 + z)^2}}.
\end{equation}

The solution of the wave equation above is specified only by: (i) the parameter $\sigma=v/U$ and (ii) the maximum value of the energy density, defined as being $\epsilon/z$ where $\epsilon$ is the wave energy. In our numerical scheme, the parameter $\epsilon$ is a normalized maximum amount of energy which we have considered equals to 1, so that when the wave amplitude $z$ goes from $z_0=2.6$ to $z(\phi=0)=0$ the energy density goes to infinite. Actually, this procedure determines the initial conditions for the integration (performed until $\Delta z=2.650$ in 24 steps, which gives reliable results considering our proposals). A canonical solution obtained for $\sigma^2=0.3$, in the domain $4\geq z > -1$ with $z_0=2.650$, using a $64\times 32$ discretized grid is shown in Fig. 2.
}
	
\begin{figure}
\centering\label{cusp}
      \includegraphics[width=0.96\textwidth]{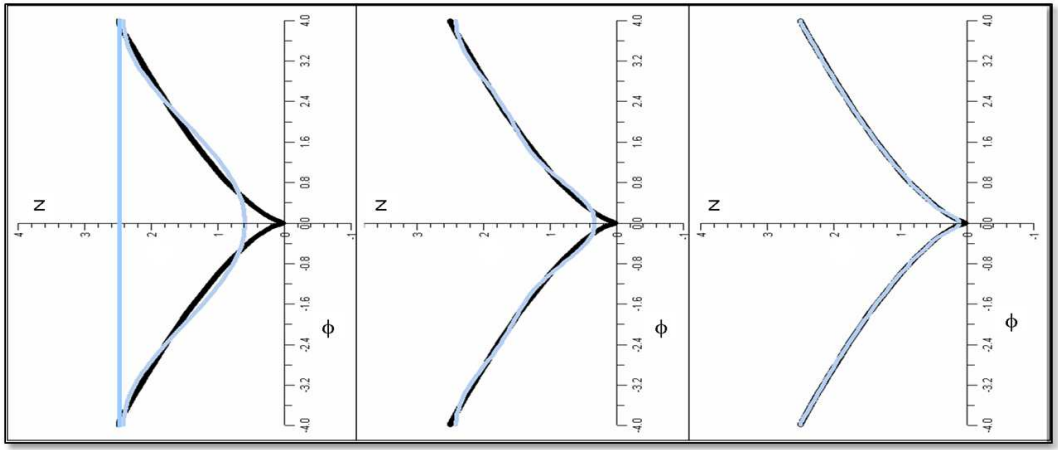}\vspace{2ex}
\caption{Profiles of the function $z=f(\phi=\Phi+U\tau)$. Snapshots at $z=2.6$ and $z=0.76$ (left panel), $z=0.38$ (middle), and $z=0.038$ (right panel). The black curve corresponds to the target solution which is the energetic cusp final state.\label{f2}}
\end{figure}

{\footnotesize
 As one sees from Fig. 3, as the parametric wave amplitude increases nonlinearity dominates the process and a cusp forms in the interval ({$0.038 > z > 0$}),  the region in which $z$ goes to zero  (the infinity energy density location). This result is important since it provides, based on the previous formalism, the concept of {\it wave-cusp parametric instability} (WCPI) to describe the physical process which we are interested. Pursuing the details of this physical process we might then to introduce the energetic cusp hypothesis.
}



\bibliographystyle{aipproc}   


\end{document}